\documentclass[twocolumn,tighten]{aastex61}
\usepackage{amssymb, amsmath,framed}

\usepackage{bm}
\expandafter\ifx\csname package@font\endcsname\relax\else
 \expandafter\expandafter
 \expandafter\usepackage
 \expandafter\expandafter
 \expandafter{\csname package@font\endcsname}
\fi
\def\bq{\begin{equation}}
\def\eq{\end{equation}}
\def\bqy{\begin{eqnarray}}
\def\eqy{\end{eqnarray}}

\begin{document}
\title{Fast Radio Bursts from Extragalactic Light Sails}

\correspondingauthor{Manasvi Lingam}
\email{manasvi@seas.harvard.edu}

\author{Manasvi Lingam}
\affiliation{Harvard John A. Paulson School of Engineering and Applied Sciences, Harvard University, Cambridge, MA 02138, USA}
\affiliation{Harvard-Smithsonian Center for Astrophysics, The Institute for Theory and Computation, 60 Garden Street, Cambridge, MA 02138, USA}

\author{Abraham Loeb}
\affiliation{Harvard-Smithsonian Center for Astrophysics, The Institute for Theory and Computation, 60 Garden Street, Cambridge, MA 02138, USA}

\begin{abstract}
We examine the possibility that Fast Radio Bursts (FRBs) originate from the activity of extragalactic civilizations. Our analysis shows that beams used for powering large light sails could yield parameters that are consistent with FRBs. The characteristic diameter of the beam emitter is estimated through a combination of energetic and engineering constraints, and both approaches intriguingly yield a similar result which is on the scale of a large rocky planet. Moreover, the optimal frequency for powering the light sail is shown to be similar to the detected FRB frequencies. These `coincidences' lend some credence to the possibility that FRBs might be artificial in origin. Other relevant quantities, such as the characteristic mass of the light sail, and the angular velocity of the beam, are also derived. By using the FRB occurrence rate, we infer upper bounds on the rate of FRBs from extragalactic civilizations in a typical galaxy. The possibility of detecting fainter signals is briefly discussed, and the wait time for an exceptionally bright FRB event in the Milky Way is estimated.
\end{abstract}

\section{Introduction} \label{SecIntro}
Ever since the first discovery of Fast Radio Bursts (FRBs) over a decade ago \citep{LBMNC}, there has been a great deal of interest in uncovering their origin. Currently, only 17 FRBs have been recorded, and a summary of their properties can be found in \citet{Pet16}.\footnote{\url{http://www.astronomy.swin.edu.au/pulsar/frbcat/}} Hypotheses put forward for FRBs range from supramassive neutron stars \citep{FR14} to gamma-ray bursts \citep{Zhang14} and stellar flares \citep{LSM14}. Regardless of their actual origin, it is now widely accepted that most FRBs are at cosmological distances \citep{Thorn13}, particularly owing to the recent localization of the repeating FRB 121102 to a dwarf galaxy at a redshift of $z \sim 0.2$ \citep{Chat17,Tend17}.

The unusually high brightness temperature of FRB sources at cosmological distances \citep{Katz16} implies that their radio emission mechanism must be coherent as known to exist in pulsars or human-made radio transmitters. Despite the diversity of explanations advanced for FRBs, the possibility that they may be of artificial origin has not been specifically investigated, apart from Section V of \citet{LuGo16}.

In this Letter, we examine the possibility that FRBs are artificial beams\footnote{Alternatively, FRBs could also be pulsar beams \citep{JKatz16}.} which have been set up as beacons, or for driving light sails. The idea that extraterrestrial civilizations may be using radio beams (manifested as dispersed pulses) is certainly not a new one, as it dates back to the pioneering paper by \citet{CM59}. This idea was extended by researchers engaged in the Search for Extraterrestrial Intelligence (SETI), accounts of which can be found in \citet{DS92} and \citet{Tart01} (see also \citealt{Sie12}). In addition to traditional, radio-based SETI, many other approaches have been advanced for detecting alien civilizations \citep[e.g][]{Dys60,ST61,Het04,BBB10,Wet14,LL17}.

In Section \ref{SecConstraints}, we show that the parameters required for generating artificial beams (to possibly power light sails) are compatible with FRB constraints. We discuss the implications and predictions in Section \ref{SecImps}, and summarize our conclusions in Section \ref{SecConc}.

\section{Compatibility of Fast Radio Bursts and Beams}\label{SecConstraints}
We start by examining whether some of the major FRB constraints are consistent with the assumption of artificial beams, and then explore the possibility that these beams may be used to power light sails.

\subsection{FRB constraints and requirements} \label{SSecCons}
We begin by denoting the distance of the beam source from the Earth by $r$. One of the primary \emph{observable} parameters for FRBs is the dispersion measure (DM), defined through a line-of-sight integral,
\begin{equation} \label{DMexp}
    \mathrm{DM} = \int_0^r n_e(s)\,ds = \bar{n}_e r,
\end{equation}
of the mean number density of free electrons $\bar{n}_e$. Ignoring contributions from the source, its host galaxy and the Milky Way, one may adopt the mean comoving electron density for the intergalactic medium (IGM), $\bar{n}_e \approx 2 \times 10^{-7}\,\mathrm{cm}^{-3}$ \citep{Planck16,FL16}. For simplicity, we assume that the FRB redshifts are $<1$, and drop the redshift factors in the context of our order-of-magnitude estimations. 

Most sources in the FRB catalog \citep{Pet16} have DM values of order hundreds of cm$^{-3}$ pc. Using equation (\ref{DMexp}), the distance can be estimated as,
\begin{equation} \label{distFRB}
r \sim 1\,\mathrm{Gpc}\,\left(\frac{\mathrm{DM}}{200\,\mathrm{cm}^{-3}\,\mathrm{pc}}\right) \left(\frac{\bar{n}_e}{2 \times 10^{-7}\,\mathrm{cm}^{-3}}\right)^{-1}.   
\end{equation}
This distance is also consistent with more accurate estimates \citep{Pet15,Pet16}.

Next, we suppose that the beam has an angular width $\theta$ and a radiated (peak) power $P$. The beam angle is expressible as some factor $\eta$ times the minimum value set by the diffraction limit,
\begin{equation} \label{AnguSize}
\theta  = \eta\, \left(\frac{c}{\nu D}\right),
\end{equation}
where $\nu$ is the frequency of the radiation, $D$ is the diameter of the beam emitter, and $\eta \geq 1$. The spectral flux density is given by,
\begin{equation} \label{SpecFluxDens}
S_\nu = \eta^{-2} \left(\frac{D}{c r}\right)^2 \alpha \varepsilon \nu P,
\end{equation}
where $\alpha = d\ln S/d\ln \nu$ is the spectral index needed for the relation $\alpha P/\nu = P_\nu$ and $\varepsilon$ is the fraction of $P$ that is emitted as FRB radiation (radiative efficiency). We have assumed that $\Delta \nu/\nu \sim 1$ and will use $\varepsilon \alpha \sim 1$ henceforth \citep{LuGo16,Katz16}, which implies that the artificial beams are \emph{broadband}. This unorthodox assumption, which is contrary to current terrestrial engineering design, will be central to our analysis, and does represent a significant conceptual challenge as to why civilizations would opt for broadband emission.\footnote{Potential reasons include: (i) photon recycling for multiple passes between the spacecraft and the planet. Each cycle yields the standard kick but one gains by the number of cycles relative to the case where the beam is dispersed after one reflection. The frequency will be Doppler shifted in each pass, and once the spacecraft reaches relativistic speeds the beam could be broadband because it will include a mix of photons with different number of passes. (ii) optimization of engineering and economic costs and functionality \citep{BBB10,Mess12}.} The mean spectrum of FRBs is observed to be Gaussian-shaped, centered on a frequency of a few GHz \citep{Law17}. While being consistent with an artificial origin, this spectral shape is puzzling for pulsar or magnetar-like sources which are expected to show power-law spectra. Inverting equation (\ref{SpecFluxDens}) yields,\footnote{We obtain Eq. (20) in \citet{LuGo16}, although the factor of $b^{-1}$ should be corrected to $b^{-2}$ in that paper.}
\begin{eqnarray} \label{Power}
P &\sim& 10^{25}\,\mathrm{erg\,s^{-1}}\, \left(\frac{r}{1\,\mathrm{Gpc}}\right)^{2}\, \left(\frac{\nu}{1\,\mathrm{GHz}}\right)^{-1}\, \left(\frac{S_\nu}{1\,\mathrm{Jy}}\right) \nonumber \\
&& \hspace{0.33 in} \times\,\eta^2\,\left(\frac{D}{3\times 10^9\,\mathrm{cm}}\right)^{-2},
\end{eqnarray}
where the value of $D$ was normalized to the size of a large rocky planet \citep{WF15} for reasons explained below.

First, let us suppose that extraterrestrial civilizations adopt the strategy of harnessing stellar energy \citep{Lubin16} to power the beams, thereby leading to
\begin{equation} \label{SolPow}
P \sim \mathcal{S} D^2 \sim 10^{25}\,\mathrm{erg\,s^{-1}}\,\left(\frac{\mathcal{S}}{\mathcal{S}_\odot}\right) \left(\frac{D}{3\times 10^9\,\mathrm{cm}}\right)^{2},
\end{equation}
where $\mathcal{S}$ is the stellar irradiance, which should be close to the solar value $\mathcal{S}_\odot$ in the habitable zone. The characteristic value of $D$ is found from (\ref{SpecFluxDens}) and (\ref{SolPow}) by substituting the typical values for all other parameters.

As the aperture efficiency is $\varepsilon$, a fraction $\left(1-\varepsilon\right)$ would be dissipated, amounting to a power per unit area of $\left(1-\varepsilon\right)P/D^2$ at the base of the emitter. If we assume that this excess heat is radiated away thermally, we get
\begin{equation} \label{HeatBal}
\frac{\left(1-\varepsilon\right) P}{D^2} = \sigma T^4,
\end{equation}
where $T$ is the surface temperature of the beamer. If the value of $T$ is too high, structural damage may follow. Hence, an upper bound on $T$ translates to a lower bound on $D$. This leads us to
\begin{equation} \label{HeatPow}
P = 10^{25}\,\mathrm{erg\,s^{-1}}\,\left(\frac{1}{1-\varepsilon}\right)\left(\frac{D}{3\times 10^9\,\mathrm{cm}}\right)^{2} \left(\frac{T}{373\,\mathrm{K}}\right)^4,
\end{equation}
and a comparison with (\ref{Power}) and (\ref{SolPow}) reveals that the same power estimate follows for $D \sim 3 \times 10^9$ cm. This represents the \emph{minimum} aperture diameter that is required to keep the system running. Note that the value of $T$ has been normalized to the boiling temperature of water, since it is widely used as a coolant in many beamer designs \citep{WNDRW}.

Thus, we have shown that the characteristic value of $D \sim 3\times10^9$ cm is obtained in two very different ways, from energetic and engineering requirements. This already constitutes a remarkable coincidence. It is rendered more unique because of a third coincidence - the value is about twice the diameter of the Earth. In other words, the beam emitter is an object akin to a planet; more precisely, it lies fairly close to the boundary between super-Earths and mini-Neptunes \citep{Rog15}. Another possibility worth considering is that the emitter could have been fashioned along the lines of the Stapledon-Dyson sphere \citep{Stap37,Dys60}.

To summarize, we assumed that: (i) the emission was broadband, (ii) stellar energy was used to power the beams, and (iii) water was used as a coolant. A subtle distinction is that (i) relies on advanced technological choices (and underpinnings), whilst (ii) and (iii) are tantamount to statements about the availability and usefulness of raw materials.

\subsection{What is the purpose of these beams?}
The preceding discussion concluded that some of the major observables for FRBs are consistent with the idea that they may be manifestations of extragalactic beams. However, this still fails to answer the important question of why they would be extant.

The first, and most immediate, possibility is that they serve the purpose of `beacons', and are thus meant to broadcast the presence of alien civilizations. But, why would a civilization want to broadcast its presence? In \citet{BBB10}, a variety of motives were considered, many of them of a sociological origin, such as a call for help, a desire to proclaim their technological achievements, etc. Although these possibilities cannot (and ought not) be ruled out, there are some inherent difficulties. They rely on complex (anthropocentric) reasons to some degree, and are thus not easily testable. Moreover, equation (\ref{Power}) demonstrates that a power of $10^{25}\,\mathrm{erg\,s^{-1}}$ is required, representing a fairly high expenditure. Hence, it seems rather implausible that this power would be expended on merely broadcasting a civilization's existence.

Instead, we consider the idea discussed in \citet{BBB10}, further elaborated in \citet{GL15} and \citet{BB16}, that these beams may power light sails. Suppose that a civilization wishes to construct a light sail capable of attaining relativistic speeds. In \citet{GL15}, it was argued that an efficient strategy for achieving the largest possible velocity for a limited acceleration value leads to
\begin{equation}
v_\mathrm{max} = \sqrt{2a_\mathrm{max}d_F},   
\end{equation}
where $v_\mathrm{max}$ and $a_\mathrm{max}$ are the maximum velocity and acceleration respectively, whilst $d_F = \nu D^2/c$ is the Fresnel distance. The above expression takes advantage of the constant beam diameter in the near-field Fresnel region (with the sail size matching $D$) out to $d_F$, where the beam enters the Fraunhofer (far-field) regime and starts to diverge with an opening angle $\theta$. In Section \ref{SSecCons}, we argued that $D$ should be normalized in units of $3 \times 10^9$ cm for a multitude of reasons; this amounts to $d_F \sim 0.1\,\mathrm{pc}$. Using this value along with the characteristic values for $v_\mathrm{max}$ and $a_\mathrm{max}$, we arrive at
\begin{equation} \label{Frequency}
\nu = 1.5\,\mathrm{GHz}\,\left(\frac{v_\mathrm{max}}{c}\right)^2 \left(\frac{a_\mathrm{max}}{1\,\mathrm{gee}}\right)^{-1} \left(\frac{D}{3\times 10^9\,\mathrm{cm}}\right)^{-2},
\end{equation}
having normalized the acceleration in the anthropic units of $1$ gee. Remarkably, the above frequency coincides with characteristic value of $1$ GHz considered thus far, which implies that the beam frequency that is optimal for powering the light sail falls within the range of FRB frequencies. Thus, it seems somewhat reasonable to hypothesize that the beams can be used to power light sails, and the ensuing implications are explored next.

\section{Discussion} \label{SecImps}
Next, we delve into some of the other consequences arising from our prior analysis.

\subsection{The angular velocity of the beam}
Hitherto, we have not discussed any of the temporal aspects of the beam. We begin by noting that FRBs are detected as pulses with a duration $\Delta t$ that is typically milliseconds. Suppose that the beam sweeps across the sky with an angular velocity $\Omega$ that is related to $\Delta t$ \citep{BB16} via
\begin{equation}
\frac{\eta c}{\nu D} = \theta = \Omega \Delta t.
\end{equation}
Alternatively, we can introduce the time period $\tau = 2\pi/\Omega$, which can be determined from the above formula, and is given by
\begin{equation}
\tau = \frac{7.3\,\mathrm{days}}{\eta}\, \left(\frac{\nu}{1\,\mathrm{GHz}}\right) \left(\frac{D}{3 \times 10^9\,\mathrm{cm}}\right) \left(\frac{\Delta t}{1\,\mathrm{ms}}\right).
\end{equation}
Thus, the beam has a characteristic angular velocity of $10^{-5}$ rad/s, viz. a time period around one week.

The derived sweep time of the beam direction reflects the spin or orbital motion of the beamer footprint relative to the receding sail (which cause the direction of the beam to change relative to the observer). Sweeping is likely to be operational during the relatively short acceleration and deceleration stages. 

\subsection{On the dimensions of the potential light sail}
The total beam power required for driving a sail of total mass $m_s$ and maximum acceleration $a_\mathrm{max}$ can be easily computed, assuming that the reflectivity is perfect \citep{Ben13,GL15}.
\begin{equation} \label{SailPow}
P = 3 \times 10^{25}\,\mathrm{erg\,s^{-1}}\,\left(\frac{m_s}{2\times10^{6}\,\mathrm{tons}}\right)\left(\frac{a_\mathrm{max}}{1\,\mathrm{gee}}\right),
\end{equation}
and the same characteristic value of $a_\mathrm{max}$ from (\ref{Frequency}) has been utilized. Note that $m_s$ has been normalized by that particular amount to ensure that equation (\ref{SailPow}) matches the other estimates, namely equations (\ref{Power}), (\ref{SolPow}) and (\ref{HeatPow}).

This implies that the mass of the sail is approximately $10^6$ tons. In deriving this estimate, we have assumed a rough equipartition of the total mass between the sail and the payload, implying that the latter is also $\sim 10^6$ tons. If the payload's density is akin to that of the International Space Station, the dimensions must be of order $100$ meters.

The mass is rather high by human standards; most estimates for light sail propulsion are two orders of magnitude lower \citep{Craw90,VJM15}. This estimate is approximately equal to the early fission-based rockets considered in the literature, which posited a total weight of up to $10^7$ tons \citep{Dys68}. Thus, if the beam was indeed used to power a spaceship, the latter would possibly have to be very large - an ``interstellar ark'' or ``world ship'' of sorts, although their typical masses ($\sim 10^{11}$ tons) are much higher \citep{HPPBR}.

\subsection{Implications for the number of advanced civilizations} \label{SSecFermiP}
Equation (\ref{distFRB}) and the DMs listed in the FRB catalog imply that the characteristic distance to FRBs is of order a few comoving Gpc \citep{Pet16} and the survey volume is of order $\sim \frac{4\pi}{3} (3~{\rm Gpc})^3\sim 100~{\rm Gpc}^3$. Since we know that there are $\sim 10^{10}$ habitable Earth-size planets in our Galaxy \citep{DC15,BC15,WF15}, and $\sim 10^{20}$ in the entire Hubble volume \citep{BP15}, it is fair to assume that there are $N_E \sim 10^{19}$ habitable Earth-size planets within a volume $\sim\,100$ Gpc$^3$. Of these, suppose that a fraction $f$ of these planets are broadcasting beams, manifested as FRBs. 

Next, note that the characteristic beam solid angle is $\theta^2 = \eta^2 10^{-16}$ steradians, based on the characteristic parameters from the previous sections and Equation (\ref{AnguSize}). Since the sky is comprised of $4\pi$ steradians, and there are $f \cdot N_E$ broadcasting planets, at any given point in time $\sim (10^{-16}/4\pi) f \eta^2 N_E$ beams are visible. Each beam is visible for $\Delta t \sim 1$ ms, which implies that approximately $10^{10}\, f \eta^2$ beams should be visible in a day. The latest estimates suggest that there are $\sim 10^4$ FRBs per day \citep{SSHC16}. If we posit that not \emph{every} FRB arises from extragalactic civilizations, then we find,
\begin{equation} \label{IntLifeFrac}
f \eta^2 \leq 10^{-6}.
\end{equation}

Since we know, by definition, that $\eta \geq 1$, we arrive at the conclusion that $f \leq 10^{-6}$. If each civilization broadcasts only a single beam, this allows us to place a bound on the number of technologically sophisticated civilizations. Using this value of $f$ in conjunction with the fact that there are $\sim 10^{10}$ habitable Earth-size planets in our Galaxy leads us to the conclusion that there are less than $10^4$ FRB-producing civilizations in a galaxy similar to our own. These civilizations must belong to the Kardashev I class \citep{Kar64} at the minimum,\footnote{Recently, extensive studies have been undertaken which place stringent constraints on the number of Kardashev III civilizations \citep{Wet14,Zet15}.} as seen from the characteristic power required in equation (\ref{Power}). Although this number is undoubtedly on the higher side, it is consistent with the earlier, more optimistic studies involving the famous Drake equation \citep{DS92}; some of the current theories have also yielded similar values \citep{Forg09,Ling16}.

We reiterate that the above value is an \emph{upper bound}. There exist at least three factors which can lower it:
\begin{itemize}
\item It is possible that the beam angle is not diffraction limited. Even a fairly modest choice of $\eta \approx 3$ can lower the value of $f_\mathrm{max}$ by an order of magnitude, as evident from (\ref{IntLifeFrac}). 
\item Not all FRBs have an artificial origin - only a fraction of them could correspond to alien activity. As an example, one may need to single out only those FRBs that repeat, such as FRB 121102 \citep{Mao15,SSH16,Chat17,Tend17}.
\item A civilization can set up more than one beam emitter. Although it may seem unlikely, this could very well happen if a civilization has progressed to the Kardashev II or III stages. 
\end{itemize}

An interesting corollary also follows: since we have assumed that FRBs are of planetary origin, the rate of FRBs is therefore set by the number of planets with advanced civilizations. This is in contrast to other models of FRBs, such as gamma-ray bursts \citep{Zhang14,DeLa16}, whose occurrence rate is determined by the formation rate of massive stars.

\subsection{Detecting FRBs of an artificial origin}
The power `leakage' from the light sails would be high, comparable to the beam power, because of broadband emission since the diffraction limit is different for each frequency, unlike the idealized case of monochromatic emission studied in \citet{GL15}.

It should be possible to distinguish between FRBs of natural and artificial (light sail) origin based on the expected shape of the pulse, as the beam sweeps by to power the light sail \citep{GL15}. More specifically, the sail would cast a moving shadow on the observed beam, thereby leading to a diffraction pattern and multiple peaks in the light curve based on the sail geometry \citep{ML16}. A series of short symmetric bursts would be observed as the beam's path intersects with the observer's line of sight \citep{GL15}. Hence, looking for similar signatures in the signal could help determine whether FRBs are powered by extragalactic civilizations (although the use of a broad range of frequencies might smear these signals).

We suggest that initiatives such as Breakthrough Listen\footnote{\url{https://breakthroughinitiatives.org/Initiative/1}} could be first directed towards the repeating FRB 121102 \citep{SSH16,SSHC16,Chat17,Tend17}. This proposition is reasonable since astrophysical explosions tend to produce single bursts, while artificial beacons can repeat.

\subsection{Looking beyond FRBs}
In our analysis thus far, we have explicitly worked with parameters that were characteristic of FRBs, such as $S_\nu \sim 1$ Jy. If all other quantities were held fixed in equation (\ref{SpecFluxDens}) except for the power which is lowered significantly, $S_\nu$ would be much smaller.

If the beam is assumed to power a light sail, equation (\ref{SailPow}) implies that the light sail's mass or its maximum acceleration should be reduced to lower the power. In turn, this implies that the spacecraft would not be capable of interstellar travel on short timescales; instead, it would be more likely to operate over \emph{interplanetary} distances. Hence, there may be a large number of interplanetary spacecrafts operating at extragalactic distances that are too faint to be detected. In contrast, such spacecrafts (and beams) \emph{within} our Galaxy are potentially detectable \citep{GL15}. 

Finally, we end our discussion with an interesting observation. There are approximately $10^9$ $L_\star$ galaxies within $100$ Gpc$^3$, and approximately $10^4$ FRBs per day, as discussed in Section \ref{SSecFermiP}. Thus, each Galaxy has a probability of $10^{-5}$ FRBs/day. Hence, an FRB emanating from our own Galaxy can be detected every $10^5\, \mathrm{days} \approx 300\, \mathrm{years}$ \citep{ML17}. A Galactic FRB at a distance of $10$-$20$ kpc would be truly spectacular since the expected value of $S_\nu$ would be $10^{10}$-$10^{11}$ Jy, and are detectable by low-cost radio receivers \citep{ML17}. This striking event could reveal everything that can be known about the true origin of FRBs, and thereby settle this FRB origin debate once and for all.

\section{Conclusions} \label{SecConc}
In this Letter, we have posited that Fast Radio Bursts are beams set up by extragalactic civilizations to potentially power light sails. 

In Section \ref{SecConstraints}, we showed that the FRB parameters were consistent with the assumption that they are artificial beams. We also demonstrated that there existed a `natural' size for the emitter which was approximately twice the diameter of the Earth. This value was obtained by adopting two contrasting estimates - the first from energy considerations, whilst the second followed from engineering constraints. Subsequently, we illustrated that the frequency needed to power the light sail was consistent with those observed for FRBs, lending further credence to our hypothesis.

Our analysis gave rise to many interesting consequences. It was shown that the payload of the light sail and the beam's characteristic period should be approximately $10^{6}$ tons and $7$ days respectively. Moreover, under certain simplifying assumptions, we derived an upper bound on the total number of intelligent civilizations in a galaxy (akin to the Milky Way). We also suggested that smaller light sails may be widely prevalent, which are presently undetectable as their spectral flux densities are too low. Using the all sky cosmological rate of FRBs, we argued that an FRB might originate within the Milky Way once every several centuries.

Although the possibility that FRBs are produced by extragalactic civilizations is more speculative than an astrophysical origin, quantifying the requirements necessary for an artificial origin serves, at the very least, the important purpose of enabling astronomers to rule it out with future data.

\acknowledgments
We thank James Benford, James Guillochon, Jonathan Katz, Zac Manchester, Dani Maoz, Andrew Siemion, Jason Wright and the referees for helpful comments. This work was partially supported by a grant from the Breakthrough Prize Foundation for the Starshot Initiative.


\end{document}